\input phyzzx
%
%
%
\newcount\lemnumber   \lemnumber=0
\newcount\thnumber   \thnumber=0
\newcount\conumber   \conumber=0

\def\myeq{{\rm \chapterlabel.\the\equanumber}}

\def\Lemma{\par\noindent\global\advance\lemnumber by 1
           {\bf Lemma\ (\chapterlabel\the\lemnumber)}}
\def\Corollary{\par\noindent\global\advance\conumber by 1
           {\bf Corollary\ (\chapterlabel\the\conumber)}}
\def\Theorem{\par\noindent\global\advance\thnumber by 1
           {\bf Theorem\ (\chapterlabel\the\thnumber)}}

%
%
\def\e{\adveq\eqno{\rm (\chapterlabel\the\equanumber)}}
\def\adveq{\global\advance\equanumber by 1}

\def\manyeq#1{\eqalign{#1}\e}

%
%
\font\tensl=cmsl10
\font\tenss=cmssq8 scaled\magstep1
\outer\def\quote{
   \begingroup\bigskip\vfill
   \def\endquote{\endgroup\eject}
    \def\par{\ifhmode\/\endgraf\fi}\obeylines
    \tenrm \let\tt=\twelvett
    \baselineskip=10pt \interlinepenalty=1000
    \leftskip=0pt plus 60pc minus \parindent \parfillskip=0pt
     \let\rm=\tenss \let\sl=\tensl \everypar{\sl}}
\def\from#1(#2){\smallskip\noindent\rm--- #1\unskip\enspace(#2)\bigskip}

\def\r#1{$\lb \rm#1 \rb$}

%
%
\def\rarrow{\rightarrow}

\def\semidirect{\mathrel{\raise0.04cm\hbox{${\scriptscriptstyle |\!}$
\hskip-0.175cm}\times}}

\def\mod{\mathop{\rm mod}\nolimits}

\def\ref#1{$^{#1}$}

\def\wiggle{\tilde}

\def\Tr{\mathop{\rm Tr}\limits}

\def\half{{1\over2}}
\def\lb{\lbrack}
\def\rb{\rbrack}

\def\diam{{\hbox{\hskip-0.02in
\raise-0.126in\hbox{$\displaystyle\bigvee$}\hskip-0.241in
\raise0.099in\hbox{ $\displaystyle{\bigwedge}$}}}}

%


\def\sqr#1#2{{\vcenter{\hrule height.#2pt
      \hbox{\vrule width.#2pt height#1pt \kern#1pt
        \vrule width.#2pt}
      \hrule height.#2pt}}}

\def\underwig#1{	
	\setbox0=\hbox{\rm \strut}
	\hbox to 0pt{$#1$\hss} \lower \ht0 \hbox{\rm \char'176}}

\def\bunderwig#1{	
	\setbox0=\hbox{\rm \strut}
	\hbox to 1.5pt{$#1$\hss} \lower 12.8pt
	\hbox{\seventeenrm \char'176}\hbox to 2pt{\hfil}}

\overfullrule=0pt
\Pubnum={}
\date{October, 2013}
\date{October, 2013}
\line{\hfil CPHT-RR095}
\titlepage
\title{On The Real Part of a Conformal Field Theory}
\author{Doron Gepner$^{1)}$ and Herve Partouche$^{2)}$} 

\address{1) Department of Particle Physics,
Weizmann Institute of Science, Rehovot, Israel.}
\address{2) Centre de Physique Theorique, Ecole Polytechnique\foot{Unite mixte du
CNRS et de l’Ecole Polytechnique, UMR 7644.}, F-91128, Palaiseau Cedex, France}
\abstract
Every conformal field theory has the symmetry of taking each field to
its adjoint.
We consider here the quotient (orbifold) conformal field theory obtained by
twisting with respect to this symmetry. A general method for computing such
quotients is developed using the coulomb gas representation. Examples
of parafermions, $SU(2)$ current algebra and the $N=2$ minimal models are
described explicitly. The partition functions and the dimensions of
the disordered
fields are given. This result is a tool for finding
new theories.
\endpage

Conformal field theories enjoy several operations which result in 
different
conformal field theories. Examples are toroidal orbifolds \REF\Orbifold{
L. Dixon, J. A. Harvey, C. Vafa and E. Witten, Nucl. Phys. B 261 (1985) 678;
Nucl. Phys. B 274 (1986) 285.}
\r\Orbifold, and
coset type models \REF\Cosets{
P. Goddard, A. Kent and D. Olive, Phys. Lett. B 152 (1985) 88;
Comm. Math. Phys. 103 (1986) 105.}
\r\Cosets. Every conformal field theory (CFT), $C$, contains
the fields
in the Hilbert space along with their conjugate $A^\dagger$. In general,
$A\neq A^\dagger$. Thus we may consider the quotient theory (abstract
orbifold),
$$\wiggle C=C/w,\qquad {\rm where}\ w(A)=A^\dagger.\e$$

From the point of view of CFT, this quotient is quite complicated since every
field in the Hilbert space transforms independently, and is not
organized by some
characters of an extended algebra. We found, however, the following
method to
compute the partition function. Many, if not all, known rational conformal
theories may be described by a system of free bosons moving on a Lorentzian
lattice with a background charge. This is called the coulomb gas
method.
By describing $C$ as such a system, the quotient by $w$ becomes a
$Z_2$ orbifold
where some of the free bosons flip sign. This on the other hand is
straightforward
to compute. Thus, we derive the partition function of $\wiggle C$ in
the cases
of $Z_k$ parafermions, $SU(2)_k$ current algebra and $k$th $N=2$
minimal model,
and leave to further work the consideration of other models.
\par
The left right symmetric $Z_k$ parafermion partition function is given by
\REF\GepQiu{D. Gepner and Z. Qiu,
Nucl. Phys. B  285 (1987) 423.}
\r\GepQiu,
$$Z={|\eta(\tau)|^2\over2}\sum_{l-m=0\mod 2} |c^l_m (\tau)|^2,\e$$
where $l=0,1,\ldots,k$ and $m$ is any integer modulo $2k$. Our
considerations
below may easily be generalized for any modular invariant, but for
simplicity
we consider only the left right symmetric one. Here $\eta(\tau)$ is the
Dedekind's eta function,
$$\eta(\tau)=e^{\pi i\tau/12} \prod_{n=1}^\infty \left(1-e^{2n\pi
i\tau}\right),
\e$$
and $c^l_m(\tau)$ is the so called string function, which is a
generating function
for the $L_0$ of the states in the representation of $SU(2)_k$ current
algebra
which have isospin $l/2$ and $J_3=m/2$,
$$c^l_m(\tau)=q^\alpha \Tr_{{\cal H}_m^l} q^{L_0},\e$$
where $$\alpha={l(l+2)\over 4(k+2)}-{m^2\over 4k}-{k\over8(k+2)},\e$$
a factor which was originally added to ensure modular covariance
\REF\KacBook{V.G.
Kac, Infinite dimensional Lie algebras (Birkhauser, Boston,
1983.}\r\KacBook,
but is
necessary to ensure that $\eta c^l_m$ is the parafermion partition
function \r\GepQiu.
Kac and Peterson \REF\KacPet{V.G. Kac and D. Peterson, Adv. Math.53
(1984) 125.}
\r\KacPet\ expressed $c^l_m$ as Hecke indefinite modular forms,
$$c^l_m(\tau)=\eta(\tau)^{-3}\sum_{-|x|<y\leq |x| \atop (x,y)\,{\rm or}\,
(\half-x,\half+y)\in({l+1\over2(k+2)},{m\over2k})+Z^2} {\rm sign}(x)
e^{2\pi i\tau[(k+2)x^2-ky^2]}.\e$$

As was noted in ref. \REF\Nem{D. Nemeschansky, Phys. Lett B 224 (1989) 121.}
\r\Nem, eq. (6) can be interpreted as the partition function of two bosons,
$\phi_1$ and $\phi_2$,
(after multiplying by eta function) moving on a Lorentzian lattice of
signature
$(1,-1)$ whose stress tensor is,
$$T(z)=-{1\over4}(\partial\phi_1)^2+{1\over4}(\partial\phi_2)^2+{i\over2\sqrt
{k+2}}\partial^2\phi_1,\e$$
where we included a background charge to get the correct central charge.
The parafermions can be written as,
$$\psi(z)=\half\left(\partial\phi_2-\sqrt{k+2\over k}\partial\phi_1\right)
\exp(i\phi_2/\sqrt k),\e$$
$$\psi^\dagger(z)=\half\left( \partial\phi_2+\sqrt{k+2\over
k}\partial\phi_1\right)
\exp(-i\phi_2/\sqrt k).\e$$
The lattice on which these bosons move is read from eq. (6) and is a
rectangular one of dimensions $(\sqrt{k+2},\sqrt k)$.

Similarly, for subsequent reference $SU(2)_k$ and the $N=2$ minimal
models can
be constructed using an additional free boson (the connection between
parafermions
and $SU(2)_k$ was described in ref. \REF\ZamFat{A.B. Zamolodchikov
and V.A. Fateev, Sov.Phys. JETP 82 (1985) 215.}\r\ZamFat, whereas the
connection with $N=2$ was
described in ref. \REF\Zam{A.B. Zamolodchikov, Sov. Phys. JETP 63 (5)
(1986) 913.}
\r\Zam),
with no background charge, $\phi_3$,
$$T(z)=-{1\over4}(\partial\phi_1)^2+{1\over4}(\partial\phi_2)^2-{1\over4}
(\partial \phi_3)^2+{i\over2\sqrt{k+2}}\partial^2\phi_1,\e$$
where $\phi_{1,2}$ move on the same lattice as before and $\phi_3$
moves on a
lattice of radius $\sqrt{k}$, for $SU(2)_k$ and on radius $2k(k+2)$ for
the $N=2$ case.

The $SU(2)_k$ currents, $J^\pm$ and $J_3$ are given by,
$$\manyeq{
J^+(z)&=\half\sqrt{k}\left(\partial\phi_2-\sqrt{k+2\over
k}\partial\phi_1\right)
\exp\left[{i\over\sqrt k}(\phi_2+\phi_3)\right],\cr
J^-(z)&=\half\sqrt{k}\left(\partial\phi_2+\sqrt{k+2\over
k}\partial\phi_1\right)
\exp\left[-{i\over\sqrt k}(\phi_2+\phi_3)\right],\cr
J^3(z)&=\sqrt k\partial\phi_3.\cr}$$
Similarly, the currents for $N=2$, $G^\pm$ and $J$, are obtained by
the same
expressions with the rescaling of $\phi_3$ by a factor of $\sqrt{k+2}$.

Now, it is clear from these expressions that the operation of taking
$A\rarrow
A^\dagger$ is equivalent to taking $\phi_2\rarrow-\phi_2$ and
$\phi_3\rarrow
-\phi_3$, i.e., it is a bosonic twist. We wish to compute the
partition function,
of the quotient by this bosonic twist. Since $\phi_{2,3}$ do not have
a background
charge, this is almost a standard orbifold. Consider the parafermion
case. We have
four sectors $Z(\delta_1,\delta_2)$, where $\delta_i=0,1 \mod 2$, describing
the path integral on the torus with boundary condition $(-1)^{\delta_1}$
($(-1)^{\delta_2}$) in the space (time) directions. $Z(0,0)$ is the
parafermionic partition function $Z$ described before, eq. (2).
$Z(1,0)$ is the
partition function
$$Z(1,0)=\Tr_{\cal H} (-1)^w q^{L_0-c/24} \bar q^{\bar L_0-c/24},\e$$
where $w(A)=A^\dagger$. $Z(1,0)$ receives contributions only from
states that
have zero momentum in the $\phi_2$ direction, since a state of momentum $p$,
$A_p$ goes to $A^\dagger_{-p}$  which is different and we can form
the pair of
states, $A_p\pm A^\dagger_{-p}$, which have the eigenvalue of $w$
which is $\pm1$,
thus giving a net contribution zero to eq. (12). For $p=0$ we have only the
contribution of the moments of $\partial\phi_2$ which gives,
$\eta(\tau)\eta(2\tau)^{-1}$ by standard arguments.
Thus, $Z(1,0)$ setting $y=0$ in eq. (12) and multiplying by the above
prefactor,
for the contribution from ${\cal H}_0^l$ sector (which imply that $l$
is odd) we
get a contribution,
$$B^l(\tau)=Y(1,0)\sum_{n=-\infty}^\infty (-1)^{nk}
e^{2(k+2)\pi i\tau(n+{l+1\over 2(k+2)})^2},\e$$
for $l$ even, and $B^l=0$ for $l$ odd,
where for the parafermions the prefactor is
$$Y(1,0)=\eta(2\tau)^{-1}.$$
Note that we changed some of the signs in eq. (12). This is required
for modular
invariance as we will see below.
The full partition function $Z(1,0)$ is given by
$$Z(1,0)=\sum_{l=0 \atop l=0\mod2}^k B_l B_l^*.\e$$

For $SU(2)$ and $N=2$ we can do the same. The partition function of
$SU(2)_k$ is
given by ref. \REF\GepWit{D. Gepner and E. Witten, Nucl. Phys. B278
(1986) 493.}
\r\GepWit,
$$Z=\sum_{l=0}^k \chi_l(\tau,0,0)\chi_l^*(\tau,0,0),\e$$
where $\chi_l(\tau,z,u)$ is the character of the affine $SU(2)_k$
representation
with isospin $l/2$,
$$\chi^l={\Theta_{l+1,k+2}-\Theta_{-l-1,k+2} \over \Theta_{1,2}-
\Theta_{-1,2}},\e$$
and where the level $m$ classical theta function is defined by,
$$\Theta_{n,m}(\tau,z,u)=e^{2\pi i u} \sum_{j\in n/(2m)+Z} e^{2\pi i
\tau m j^2+
2\pi i j z},\e$$
where $n$ is defined modulo $2m$.

For the $N=2$ partition function we have \REF\Gep{D. Gepner, Nucl. Phys.
B296 (1988) 757.}\r\Gep,
$$Z=\sum_{l,m,s} \left| \chi^{l(s)}_m(\tau,0,0)\right|^2,\e$$
where the sum is over $l=0,1,\ldots,k$ and $m$ and $s$ modulo
$2(k+2)$ and $4$
respectively. $s=0,2$ in the NS sector, and $s=1,3$ in the Ramond sector.
The $\chi^{l(s)}_m(\tau,z,u)$ are the characters of the $k$th $N=2$
minimal model
and are given by,
$$\chi^{l(s)}_m(\tau,z,u)=\sum_{j\mod k} c^l_{m+4j-s}\Theta_{2m+(4j-s)(k+2),
2k(k+2)}(\tau,2kz,u).\e$$

Looking on the expressions for the characters eqs. (19), we see that
again this is
just the appropriate partition function for the appropriate system of
bosons.
Thus we get the contribution of the $l$ representation by simply setting the
momentum in the $\phi_{2,3}$ directions to zero. We get the same
answer as for
the parafermions, eq. (13), with only a different prefactor to account
for the
extra boson,
$$Y(1,0)=\eta(\tau)\eta(2\tau)^{-1}.\e$$
It is the same prefactor for $SU(2)$ and $N=2$.

We can get the partition function for the twisted sectors $Z(0,1)$
and $Z(1,1)$
by modular transformations. We have $Z(1,0)(-1/\tau)=Z(0,1)(\tau)$ and
$Z(1,1)(\tau)=Z(0,1)(\tau+1)$. For consistency it is required that
all partition
functions would be invariant under $\tau\rarrow \tau+2$ and
$Z(1,1)(-1/\tau)=
Z(1,1)(\tau)$. The total partition function is,
$$Z=\half\sum_{\delta_1,\delta_2=0,1} Z(\delta_1,\delta_2).\e$$
The partition function for the untwisted sector, namely fields
invariant under
$A\rarrow A^\dagger$, is $\half(Z(0,0)+Z(1,0))$. Similarly, the
partition function
for the twisted sector, disorder fields, is $\half(Z(0,1)+Z(1,1))$.

We can write the sum in the expression for $B_l$, eq. (13) as a theta
function,
$$B_l/Y(1,0)=\Theta_{l+1,k+2}(\tau,k/2,0),\e$$
where we used the definition, eq. (17).
The theta functions transform nicely under modular transformations
\r\KacBook,
$$\Theta_{n,m}\left(-{1\over\tau},{z\over\tau},u-{z^2\over2\tau}\right)=
{1\over\sqrt{2m}}(-i\tau)^\half\sum_{l\mod2m} e^{-\pi
iln/m}\Theta_{l,m}(\tau,z,u).
\e$$
Using this, we find the partition function $Z(0,1)$ by calculating
$Z(1,0)(-1/
\tau)$. We find,
$$Z(0,1)=|Y(0,1)|^2\sum_{\gamma,\bar\gamma \mod 2(k+2)\atop
\gamma-\bar\gamma
=0\mod k+2} e^{\pi i(\gamma-\bar\gamma)/(k+2)} \Theta_{\gamma+{p\over2},k+2}
(\tau,0,0) \Theta_{\bar\gamma+{p\over2},k+2}(\tau,0,0)^*,\e$$
where $p=0$ or $p=1$, $p=k\mod2$.
$$D_\gamma=\Theta_{\gamma+{p\over2},k+2}-\Theta_{\gamma+{p\over2}+k+2,k+2}.\e$$
are the characters of
the twisted sector, with respect to the appropriate extended chiral algebra.
The fields there differ by integer dimensions. The lowest dimension field in
$D_\gamma$ (`primary fields') are the disorder fields $A_\gamma$ of
dimension,
$$\Delta_\gamma={(\gamma+p/2)^2\over 4(k+2)}+s+c/24,\e$$
where $s=0$ for $SU(2)$ and $N=2$, and $s=-1/48$ for parafermions,
$c$ is the
central charge and $\gamma$ is defined modulo $k+2$. Actually, for parafermions,
the dimensions of these disorder fields were already calculated by Zamolodchikov,
\r\Zam\ and
our results agree, in this case. The novelty here is that we obtained the characters,
as well.

In terms of these, the partition function $Z(0,1)$ is left--right symmetric.
The prefactor, $Y(0,1)$ comes from the modular transform of the $Y(1,0)$. We
find,
$${Y(1,0)={\sqrt2\over\eta(\tau/2)},\qquad {\rm for\
parafermions},}\e$$ $$
Y(1,0)={2\eta(\tau)\over \eta(\tau/2)^{-2}},
\qquad{\rm for\ SU(2)\ and\ N=2},\e$$
where we used the well know transformation properties of the eta
function, e.g.
\r\KacBook.

\ack
This work is based on a draft written in CERN in September, 1997. We thank CERN
for the kind hospitality. We publish this work now since we think it might be
useful. In particular in ref. \REF\GG{A. Genish and D. Gepner, in preparation.}\r\GG\
we use the above results to calculate the string functions of $E_6$ level $2$. 
\refout
\bye